\documentclass[english,prd,aps,10pt,showkeys,nofootinbib,notitlepage]{revtex4-1}
\usepackage[T1]{fontenc}
\usepackage[latin9]{inputenc}
\usepackage{babel}
\usepackage{amssymb}
\usepackage{graphicx}
\usepackage{esint}
\usepackage[unicode=true,pdfusetitle,
 bookmarks=true,bookmarksnumbered=false,bookmarksopen=false,
 breaklinks=false,pdfborder={0 0 1},backref=false,colorlinks=false]
 {hyperref}

\makeatletter
\usepackage{babel}
\usepackage{babel}

\usepackage{babel}

\makeatother

\begin{document}

\title{Digluon contribution to $J/\psi$ production}

\author{Iván Schmidt, Marat Siddikov}

\affiliation{Departamento de Física, Universidad Técnica Federico Santa María,~\\
 y Centro Científico - Tecnológico de Valparaíso, Casilla 110-V, Valparaíso,
Chile}
\begin{abstract}
In this paper we study the contribution of the double parton distributions
of gluons to the charmonium production. Despite being suppressed in
the heavy quark mass limit, numerically this contribution gives a
sizeable correction to the leading order $k_{T}$ factorization result
in LHC kinematics due to enhancement of gluonic densities in the small
Bjorken $x_{B}$ limit. This contribution is not suppressed at large
$J/\psi$ momenta $p_{T}$ and thus presents one of the complementary
mechanisms of charmonia production in this kinematics. 
\end{abstract}
\maketitle

\section{Introduction}

The description of the charmonium hadroproduction remains one of the
long-standing puzzles almost since its discovery. The large mass $m_{c}$
of the charm quark inspired applications of perturbative methods and
consideration in the formal limit of infinitely heavy quark mass~\cite{Korner:1991kf}.
However, in reality the coupling $\alpha_{s}\left(m_{c}\right)\sim1/3$
is not very small, so potentially some mechanisms suppressed in the
large-$m_{c}$ limit, numerically might give a sizeable contribution. 

\begin{figure}
\includegraphics[scale=0.6]{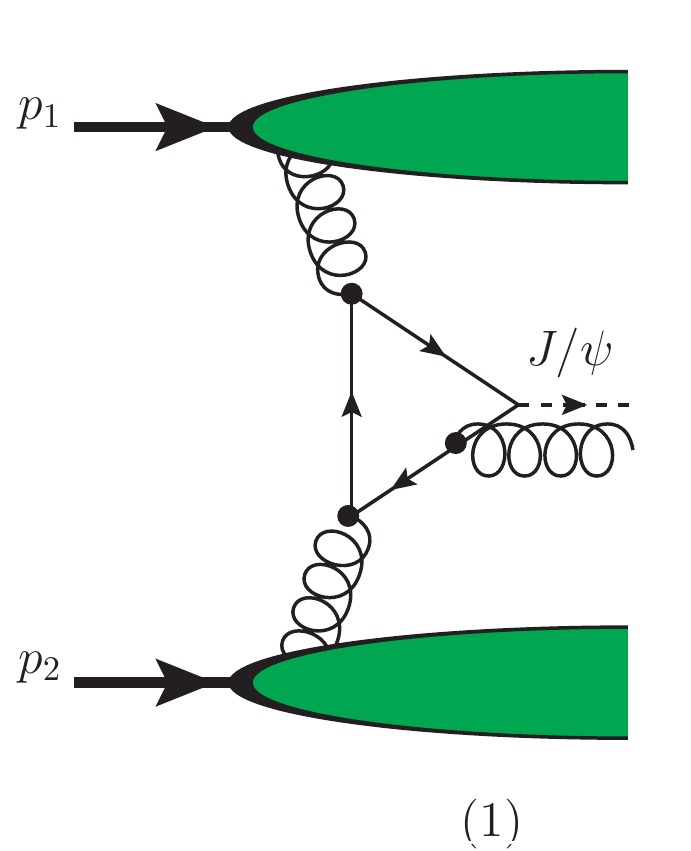}\includegraphics[scale=0.6]{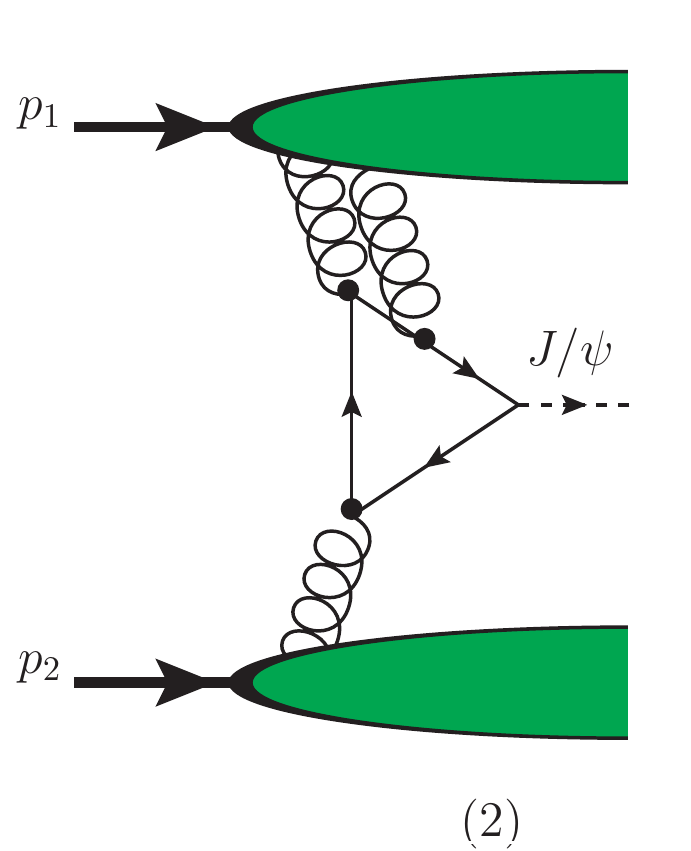}\includegraphics[scale=0.6]{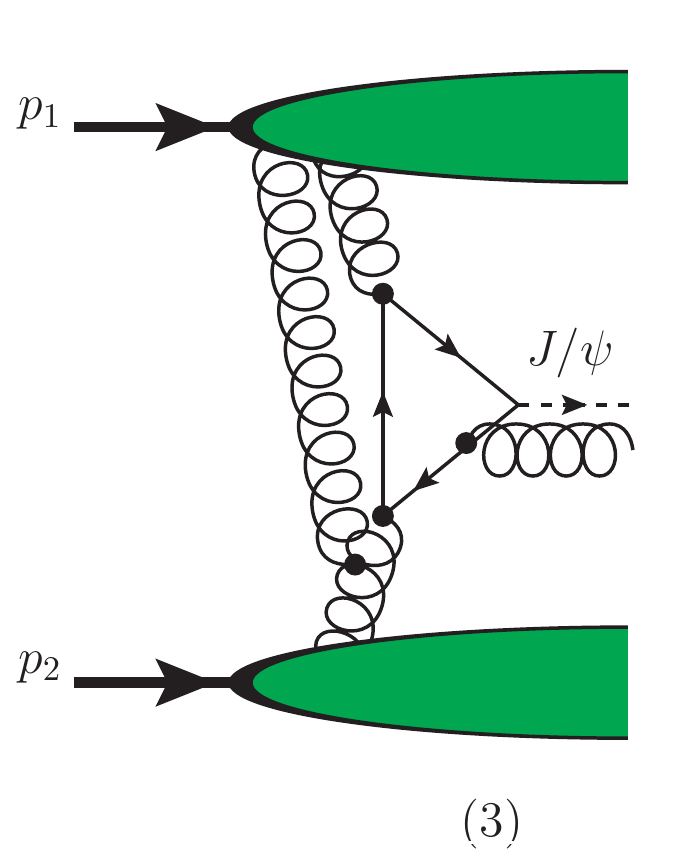}\caption{\label{fig:ProcessListCSM}Diagram (1): A conventional Color Singlet
Model (CSM) gluon-gluon fusion mechanism of $J/\psi$ production.
In our evaluations we also take into account feed-down contributions
from $\chi_{c}$ and $\psi(2S)$ decays, whose production amplitudes
have similar topology  (no gluon emission from quark loop in case of $\chi_{c}$). Diagram (2): a higher twist mechanism suggested
in~\cite{Motyka:2015kta}. Diagram (3): Example of the subprocess
in which digluons may produce the same final state as CSM process
(this paper, see section~\ref{sec:Evaluation} for details). The
two-gluon contribution may stem from either hadron. In all three diagrams
summation over all permutations of gluons in heavy quark loop is implied.}
\end{figure}

The Color Singlet Model (CSM) of charmonia production~\cite{Chang:1979nn,Baier:1981uk,Berger:1980ni}
assumes that the dominant mechanism is the gluon-gluon fusion supplemented
by emission of additional gluon, as shown in the diagram 1 of the
Figure~\ref{fig:ProcessListCSM}. Early evaluations in the collinear
factorization framework led to incorrect results at large transverse
momenta $p_{T}$ of charmonia and premature conclusions about the
inability of CSM to describe the experimental data. The failure of
the expansion over $\alpha_{s}$ due to milder suppression of higher
order terms at large $p_{T}$~\cite{Brodsky:2009cf,Artoisenet:2008fc}
and co-production of additional quark pairs~\cite{Artoisenet:2007xi,Karpishkov:2017kph}
motivated introduction of the phenomenological color octet contributions~\cite{Cho:1995vh,Cho:1995ce}.
The modern NRQCD formulation~\cite{Bodwin:1994jh,Maltoni:1997pt,Brambilla:2008zg,Feng:2015cba,Brambilla:2010cs}
constructs a systematic expansion over the Nonpertrubative Matrix
Elements (NMEs) of different charmonia states which can be extracted
from fits of experimental data. However, at present extracted matrix
elements depend significantly on the technical details of the fit~\cite{Feng:2015cba},
which sheds doubts on the universality of extracted NMEs. At the same
time, it was suggested that the results of the CSM evaluated in the
$k_{T}$-factorization framework ($k_{T}$-CSM for short) might agree
better with experimental data at large $p_{T}$ if the feed-down contributions
from $\chi_{c}$ and $\psi(2S)$ decays are taken into account~\cite{Baranov:2002cf,Kniehl:2006sk,Kniehl:2006vm,Baranov:2007dw,Baranov:2011ib,Saleev:2012hi}.
Inclusion of color octet contributions in $k_{T}$-CSM framework improves
agreement with data~\cite{Baranov:2016clx}. However, the uncertainty
of the unintegrated parton distribution function (uPDF) is large in
this kinematics, and for this reason situation with NRQCD contributions
still remains ambiguous~\cite{Baranov:2016clx,Baranov:2015laa}.
It was suggested that at large $p_{T}$, a sizeable contribution might
come from other mechanisms, like for example gluon fragmentation into
$J/\psi$~\cite{Bodwin:2012xc,Bodwin:2014gia,Braaten:1994xb,Braaten:1995cj}.

In the aforementioned analysis it was not taken into account that
in the small Bjorken-$x_{B}$ limit, as we approach saturation regime,
the gluon densities grow rapidly, and more than one gluon from each
hadron might interact with heavy quarks. In this paper we will focus
on the first correction, which probes the Double Parton Distribution
Function (DPDF) of gluon. According to recent theoretical~\cite{Diehl:2011yj,Rinaldi:2013vpa,Diehl:2014vaa,GolecBiernat:2015aza,Rinaldi:2016mlk}
and experimental~\cite{Khachatryan:2014iia,Aaboud:2016dea,Aaboud:2016fzt,Aad:2011sp,Aad:2013bjm,Abazov:2014fha,Abazov:2015nnn,Abe:1997bp,Chatrchyan:2013xxa}
studies, these objects might have rich internal structure due to possible
correlation between partons~\cite{Calucci:2010wg}, and in view of
various sum rules which the DPDFs should satisfy~\cite{GolecBiernat:2015aza}. 

The DPDFs are usually studied in the double parton scattering (DPS)~\cite{Diehl:2011yj,Diehl:2013mla,Diehl:2014vaa,Baranov:2011ch,Diehl:2017wew}
and double Drell-Yann processes~\cite{Diehl:2015bca}. However, the
DPDFs might also contribute to the \emph{single} hadron production,
which is usually interpreted as being due to single-gluon distributions
only. In case of the charmonium production, as was noticed in~\cite{Motyka:2015kta},
the DPDFs might contribute already in the same order over $\mathcal{O}\left(\alpha_{s}\right)$,
as shown in the diagram 2 of the Figure~\ref{fig:ProcessListCSM}.
The relative contribution of the DPDF-induced process is growing with
energy and in the LHC kinematics gives a sizeable contribution, up
to twenty per cent of the theoretical prediction for the prompt $J/\psi$
hadroproduction. At large momenta this contribution is suppressed
due to additional convolution of the third gluon with $k_{T}$-dependent
gluon PDF. In this paper we suggest another mechanism, with emission
of additional gluon, as shown in the diagram 3 of the Figure~\ref{fig:ProcessListCSM}.
Formally, the cross-section of this process is suppressed as $\mathcal{O}\left(\alpha_{s}\right)$
compared to that of the diagram 1, however, as we will see below,
it gives a sizeable contribution, on par with contribution of the
diagram 2. In contrast to mechanism of~\cite{Motyka:2015kta}, our
contribution is not suppressed in the large-$p_{T}$ kinematics, and
for this reason should be taken into account in comparison with experimental
data. If one or both hadrons are polarized, the interference with
leading order diagram gives rise to transverse spin asymmetries, which
have been studied in detail theoreticaly~\cite{Yuan:2008it,Yuan:2008vn,Kang:2008ih}
and experimentally~\cite{Adare:2010bd}. In this paper we will focus
on the case of unpolarized protons for which the interference term
does not contribute. 

The paper is structured as follows. In the Section~\ref{sec:Evaluation}
we discuss the framework used for evaluations. In the Section~\ref{sec:Parametrization}
we introduce the paramertrizations of gluon PDFs and DPDFs used for
our estimates. In Section~\ref{sec:Results} we present our numerical
results and finally in Section~\ref{sec:Conclusions} we draw conclusions.

\section{Evaluation of the amplitudes}

\label{sec:Evaluation}The cross-section of the charmonium production
in the $k_{T}$ factorization~ framework reads as~\cite{Baranov:2002cf,Kniehl:2006sk,Kniehl:2006vm,Baranov:2007dw,Baranov:2011ib,Saleev:2012hi}
\begin{equation}
d\sigma=\frac{\alpha_{s}^{3}\left(\mu\right)}{512\pi^{4}\,\hat{s}^{2}}\sum_{{\rm polarization}}\sum_{{\rm spin}}\sum_{{\rm color}}\left|\mathcal{M}_{gg\to gJ/\psi}\left(\hat{s},\,\hat{t}\right)\right|^{2}\mathcal{F}\left(x_{1},k_{1\perp}\right)\mathcal{F}\left(x_{2},k_{2\perp}\right)\,d^{2}k_{1\perp}d^{2}k_{2\perp}dy\,d^{2}p_{T}\,dy_{g}\label{eq:CSM}
\end{equation}
where we introduced the shorthand notation $\hat{s}=x_{1}x_{2}s$
, the variables $(y,\,p_{T})$ are rapidity and transverse momentum
of produced charmonium, $\left(y_{g},\,k_{g\perp}\right)$ are the
rapidity and transverse momentum of the emitted gluon, $\left(x_{i},\,k_{i\perp}\right)$
are the light-cone fractions and transverse momenta of the incident
gluons, with 
\begin{eqnarray}
x_{1,2} & = & \frac{\sqrt{M_{J/\psi}^{2}+p_{T}^{2}}}{\sqrt{s}}e^{\pm y}+\frac{k_{g\perp}}{\sqrt{s}}e^{\pm y_{g}},\\
\vec{k}_{g\perp} & = & \vec{p}_{T}-\vec{k}_{1\perp}-\vec{k}_{2\perp}.
\end{eqnarray}
$\mathcal{F}\left(x_{i},\,k_{i\perp}\right)$ in~(\ref{eq:CSM})
are the unintegrated gluon parton distributions (uPDFs). The parton
level amplitude $\mathcal{M}_{gg\to gJ/\psi}$ in~(\ref{eq:CSM})
is given by a sum of diagrams with all possible permutations of gluon
vertices in heavy quark loop in diagram~1 of Figure~(\ref{fig:ProcessListCSM}).
We fix the renormalization scale $\mu$ as $\mu=\sqrt{M_{J/\psi}^{2}+p_{T}^{2}}$.
For the $J/\psi$ vertex the standard approximation is to neglect
the internal motion of the quarks (formally $\mathcal{O}\left(\alpha_{s}(m_{c})\right)$
effect) and use~\cite{Chang:1979nn,Baier:1981uk,Berger:1980ni} 
\begin{equation}
\hat{J}\left(^{3}S_{1}\right)=\frac{g\,\hat{\epsilon}\left(S_{J/\psi}\right)\left(\hat{p}+m_{c}\right)}{2}\label{eq:JPsi_vertex}
\end{equation}
where $\epsilon_{J/\psi}$ is the polarization vector of $J/\psi$
and the normalization constant $g$ is fixed from the leptonic decay
width $\Gamma_{J/\psi\to e^{+}e^{-}}$, 
\begin{equation}
g=\sqrt{\frac{3m_{J/\psi}\Gamma_{J/\psi\to e^{+}e^{-}}}{16\pi\alpha_{em}^{2}Q_{c}^{2}}},\quad Q_{c}=\frac{2}{3}.
\end{equation}
For gluon polarization vectors we used the light-cone gauge $A^{+}=0$,
in which the parton distributions have a simple probabilistic interpretation.
The evaluation of the Feynman diagrams is straightforward in the $k_{T}$
factorization framework and was done with the help of \emph{FeynCalc}
package~\cite{FC-2,FC-1}.  The code for evaluation of the cross-section~(\ref{eq:CSM})
is available on demand. 

\begin{figure}
\includegraphics[scale=0.5]{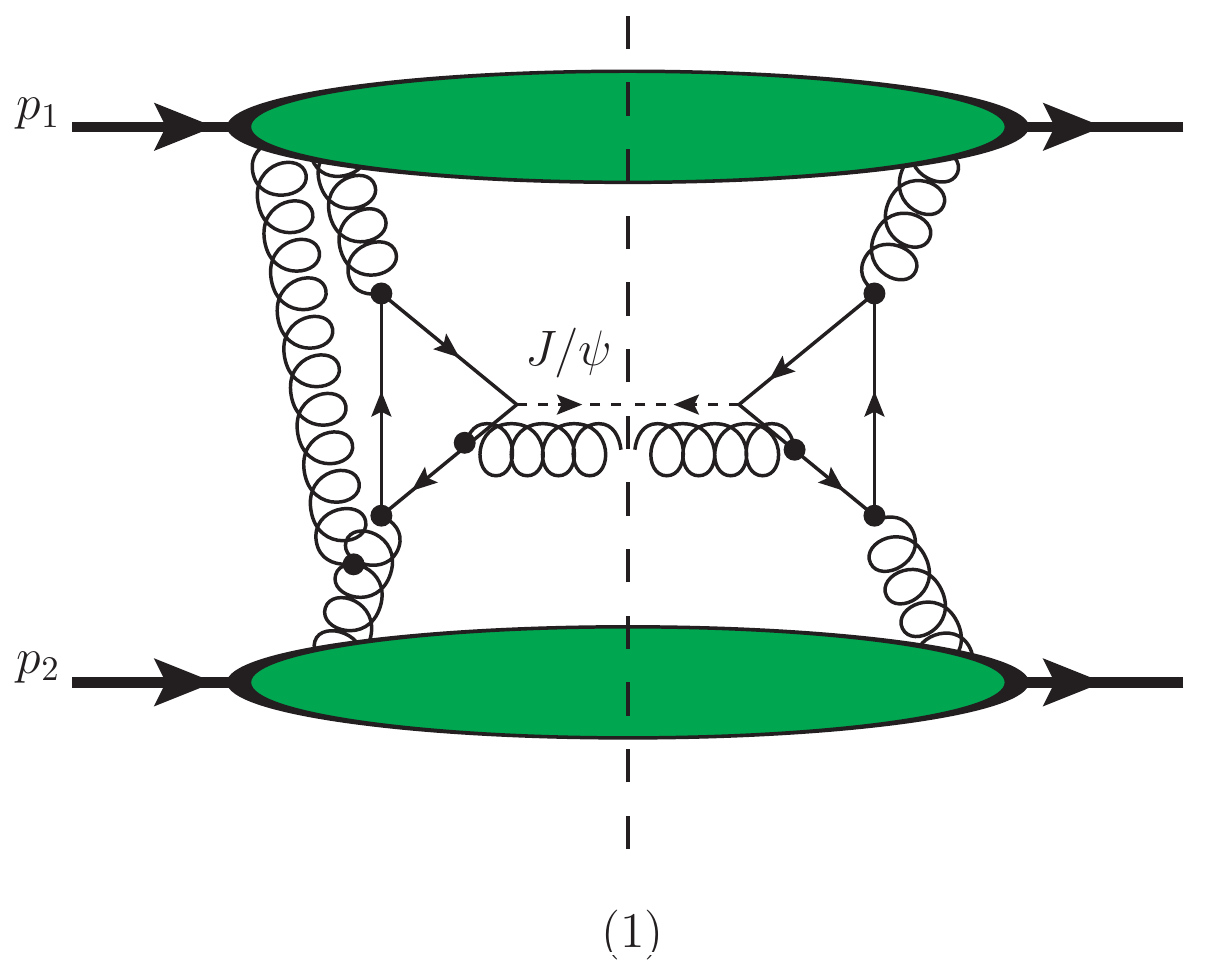}\includegraphics[scale=0.5]{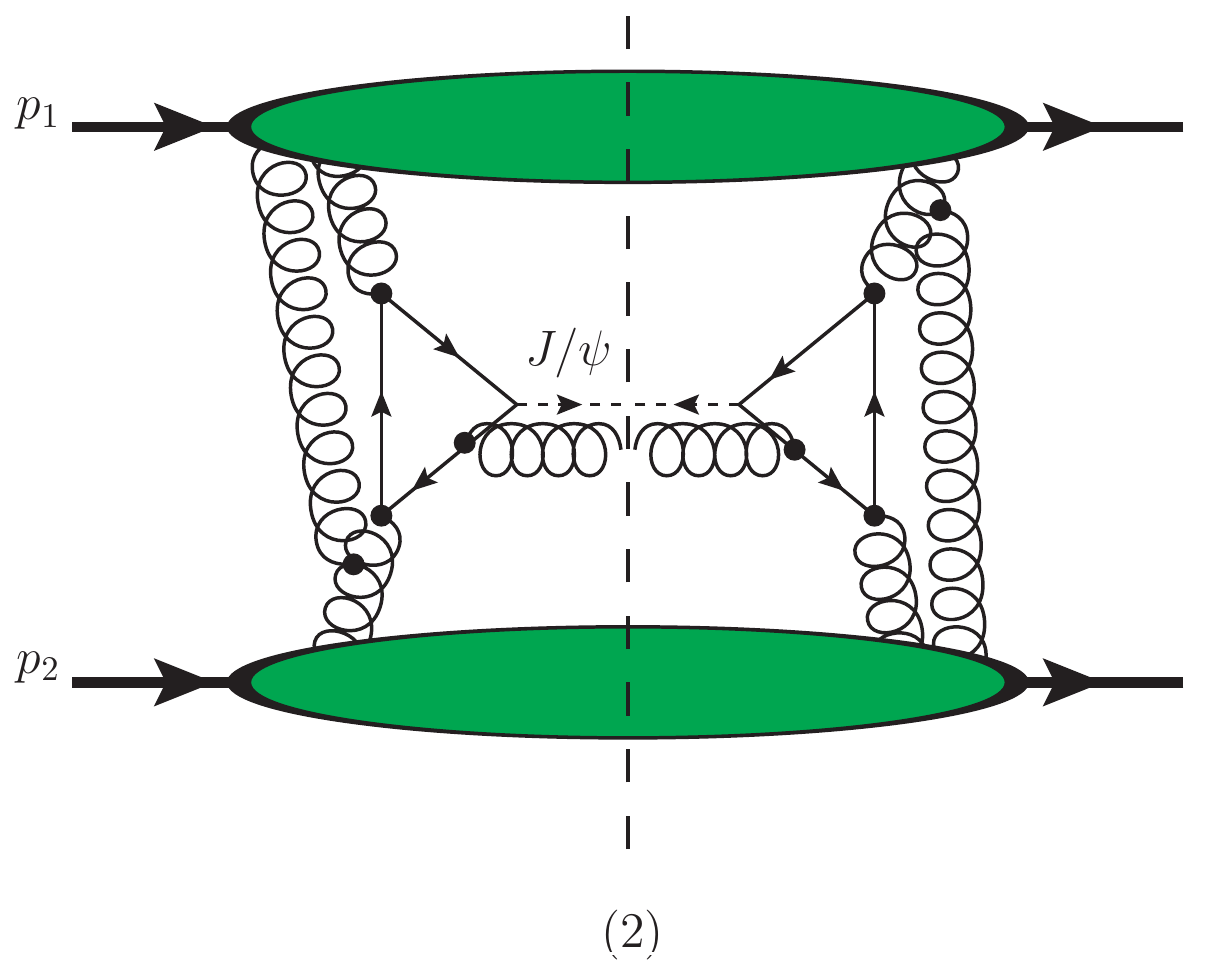}\includegraphics[scale=0.5]{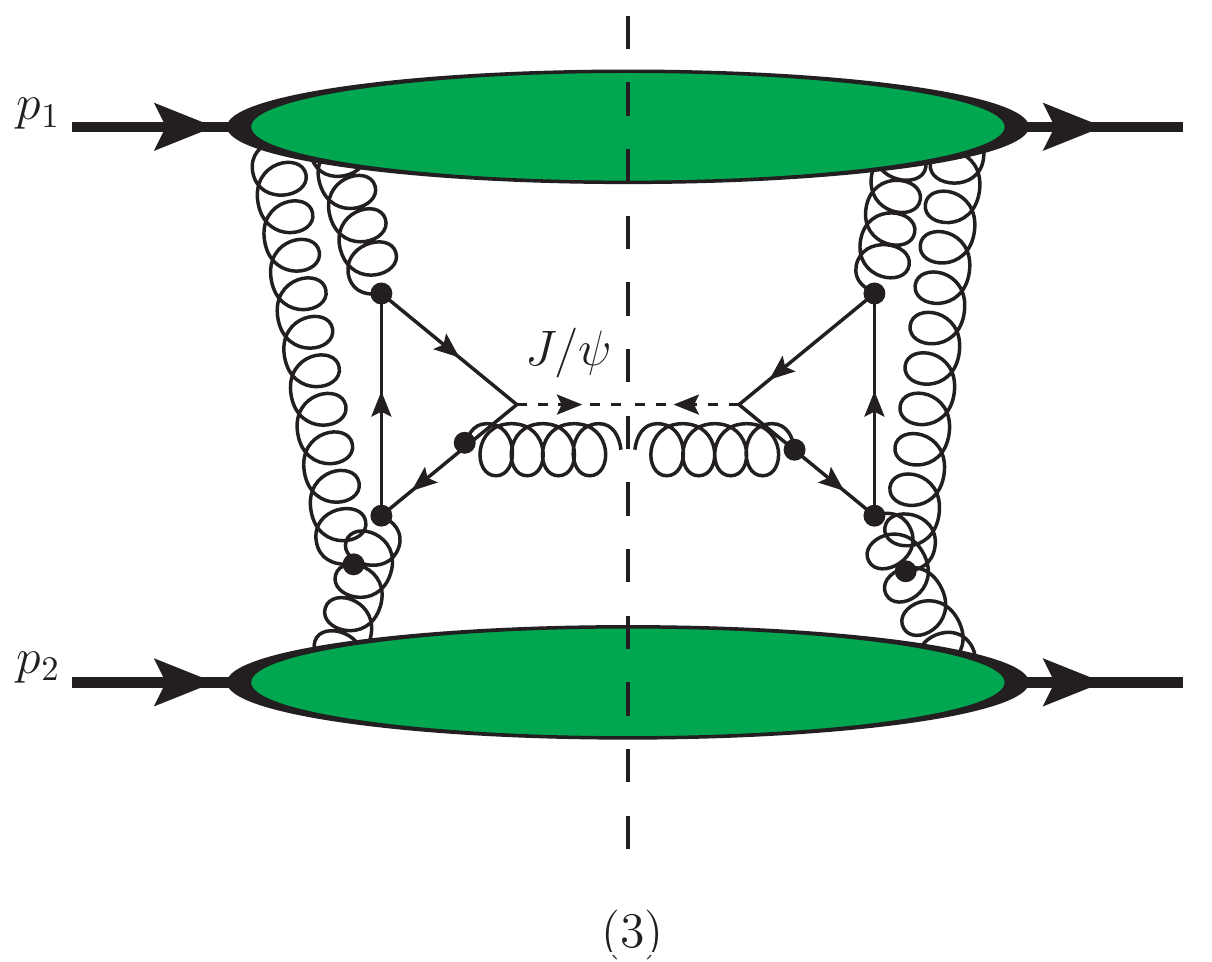}\caption{\label{fig:InterferencesLO}Diagram (1): Interference of LO and NLO
correlators which contributes only when upper hadron is polarized
and leads to charmonium spin asymmetry studied in PHENIX~\cite{Adare:2010bd}.
Diagram (2): Contribution which probes three-gluon correlators in
both hadrons and contributes only if both hadrons are polarized. Diagram
(3): Contribution from gluon DPDFs which gives nonzero result even
if both incident hadrons are not polarized. In all diagrams summation
over all permutations of gluon vertices in quark loops is implied.}
\end{figure}

The process which we study in this paper has the same final state
as the CSM mechanism and may interfere with it, as shown in the diagram
1 of the Figure~\ref{fig:InterferencesLO}. As was discussed in detail
in~\cite{Yuan:2008it,Yuan:2008vn,Kang:2008ih}, the interference
contributes only if one of the incident hadrons is transversely polarized
and leads to transverse spin asymmetry sensitive to the three-gluon
correlators suggested in~\cite{Ji:1992eu}. This asymmetry has been
measured by PHENIX collaboration~\cite{Adare:2010bd}, and very small
value compatible with zero implies that the three-gluon correlators
are negligible. For the same reason we will omit the interference
diagrams shown in the Figure~\ref{fig:InterferencesLO}: they might
contribute only if both hadrons are polarized. For the unpolarized
protons, digluons should stem from the same hadron in the amplitude
and its conjugate, as shown in the diagram 3 of the Figure~\ref{fig:InterferencesLO}.
The diagram with digluon stemming from the lower proton differs from
the diagram 3 in Figure~\ref{fig:InterferencesLO} only by inversion
of sign of rapidity $y$ of $J/\psi$, so the final result has a symmetric
form 
\begin{equation}
d\sigma_{J/\psi}(y)=d\sigma_{gg+g\to J/\psi\,g}(y)+d\sigma_{gg+g\to J/\psi\,g}(-y),
\end{equation}

\begin{figure}
\includegraphics[scale=0.6]{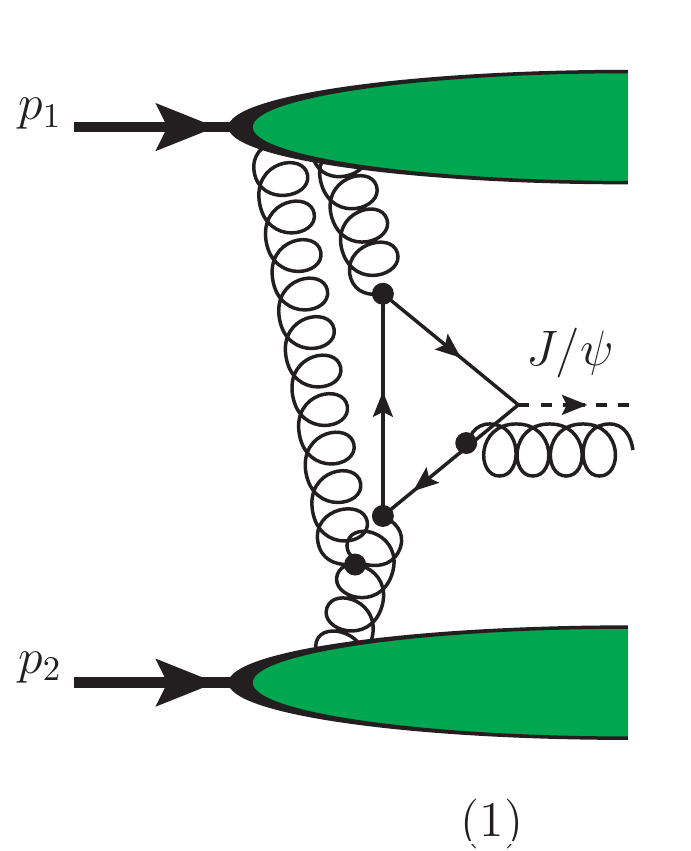}\includegraphics[scale=0.6]{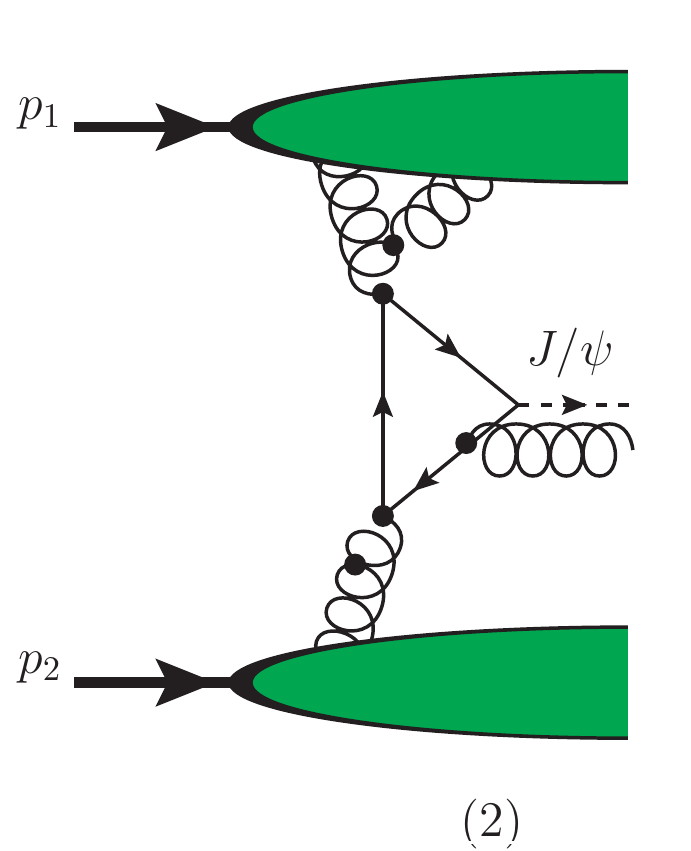}\includegraphics[scale=0.6]{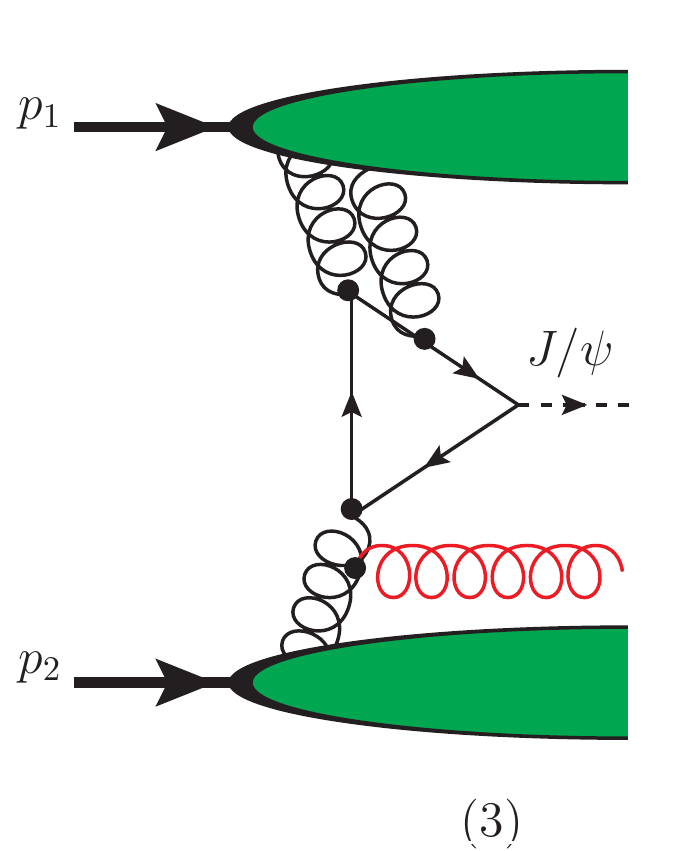}\includegraphics[scale=0.6]{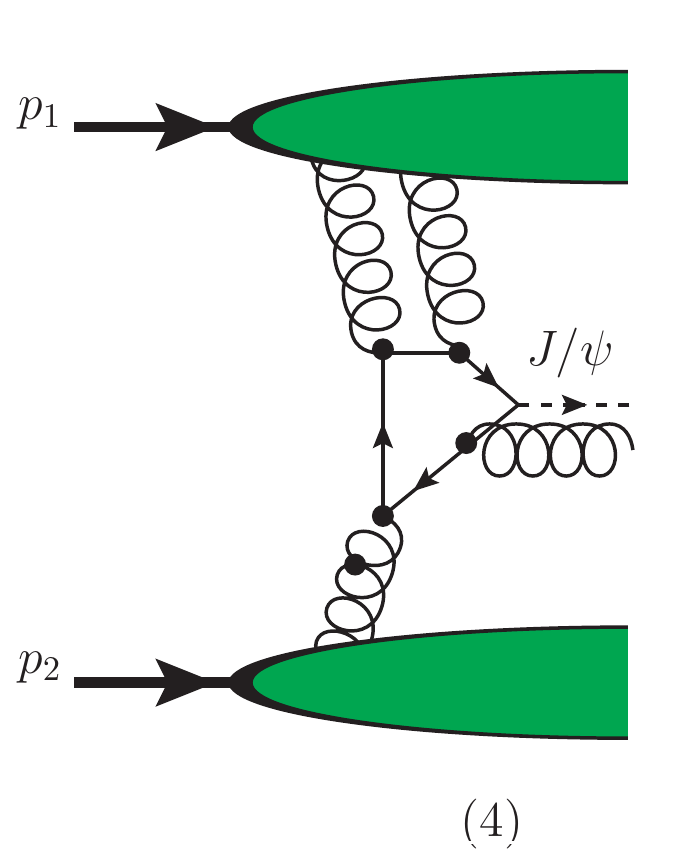}\caption{\label{fig:Gauge}Diagram (1): A digluon correction to the conventional
Color Singlet Model (CSM) $J/\psi$ production. Diagram (2): a contribution
to the gluon PDF from digluons which contributes in the same order.
Diagram (3): radiative correction to the process suggested in~\cite{Motyka:2015kta}.
Sum of diagrams with emission from any of three $t$-channel gluons
is assumed. Diagram (4): process without three-gluon vertex, exists
due to nontrivial color group structure. In all diagrams summation
over all permutations of gluon vertices in quark loop are implied.}
\end{figure}

\begin{figure}
\includegraphics[scale=2]{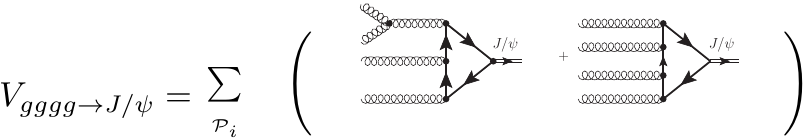}\caption{\label{fig:GluonVertex}The sum of the hard coefficient functions
of the diagrams in the Figure~\ref{fig:GluonVertex} effectively
correspond to four gluon-$J/\psi$ vertex $V_{gggg\to J/\psi}$. Summation
over all possible permutations $\mathcal{P}_{i}$ of the four gluons
is implied.}
\end{figure}

\begin{figure}
\includegraphics[scale=0.6]{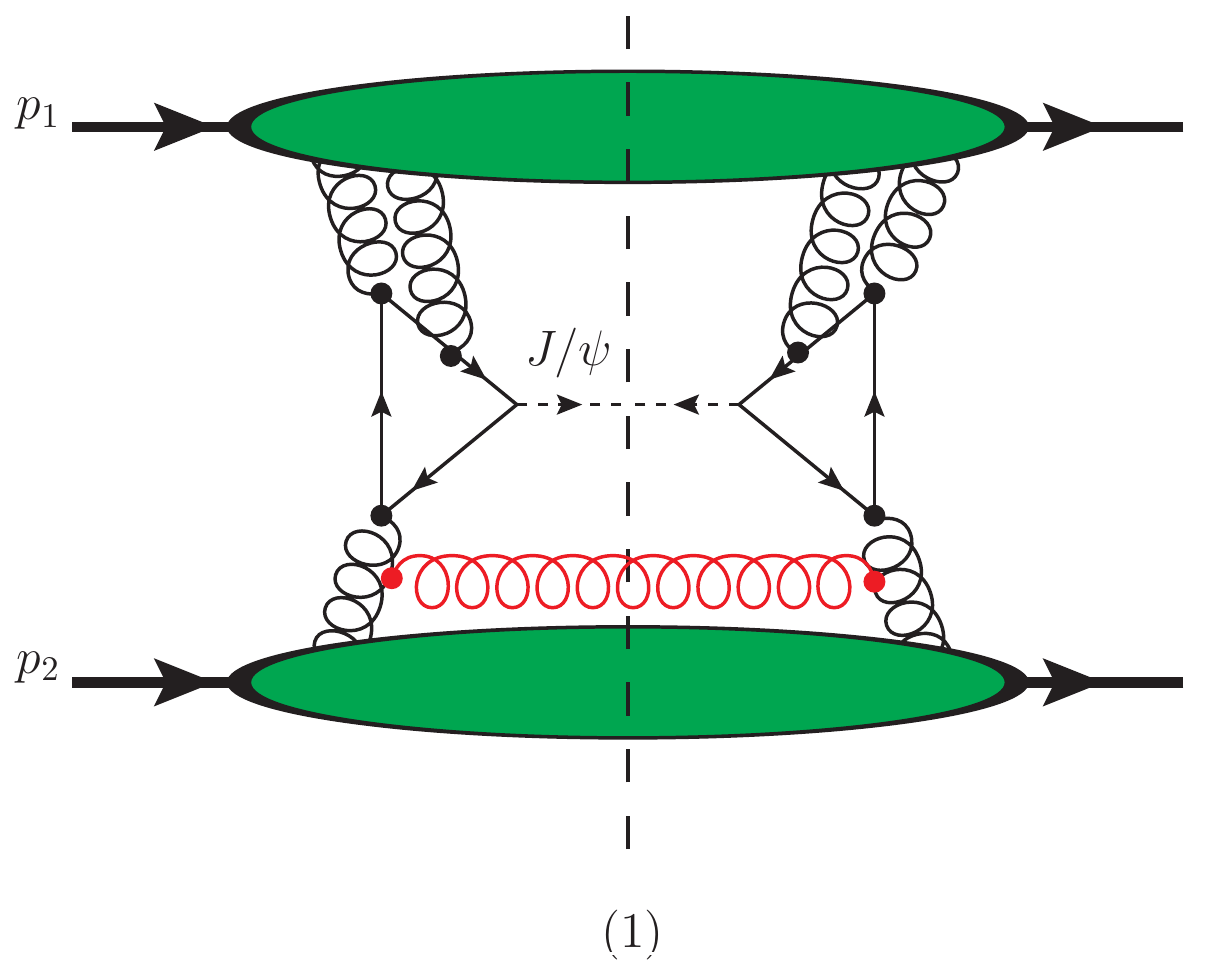}\includegraphics[scale=0.6]{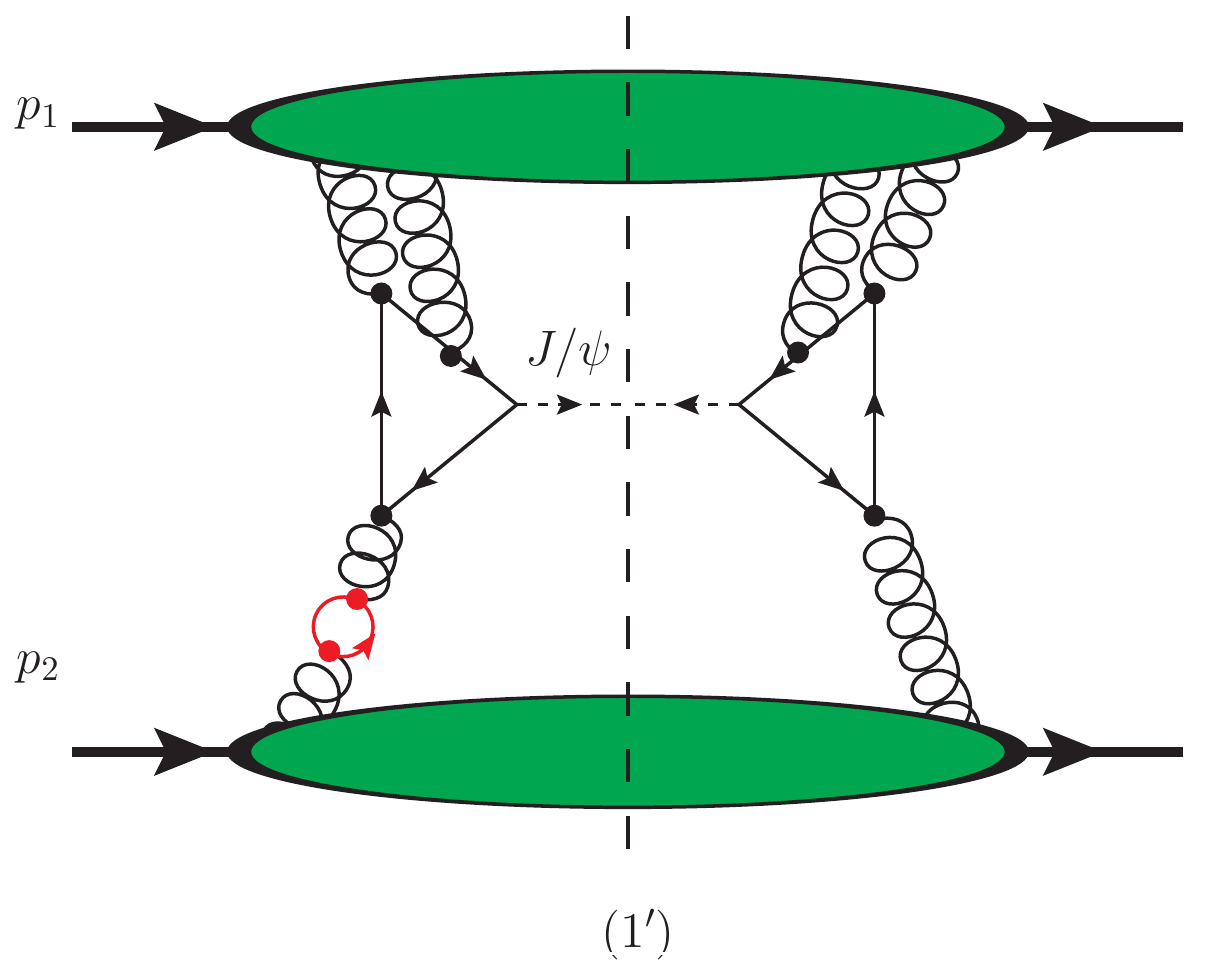}\\
 \includegraphics[scale=0.6]{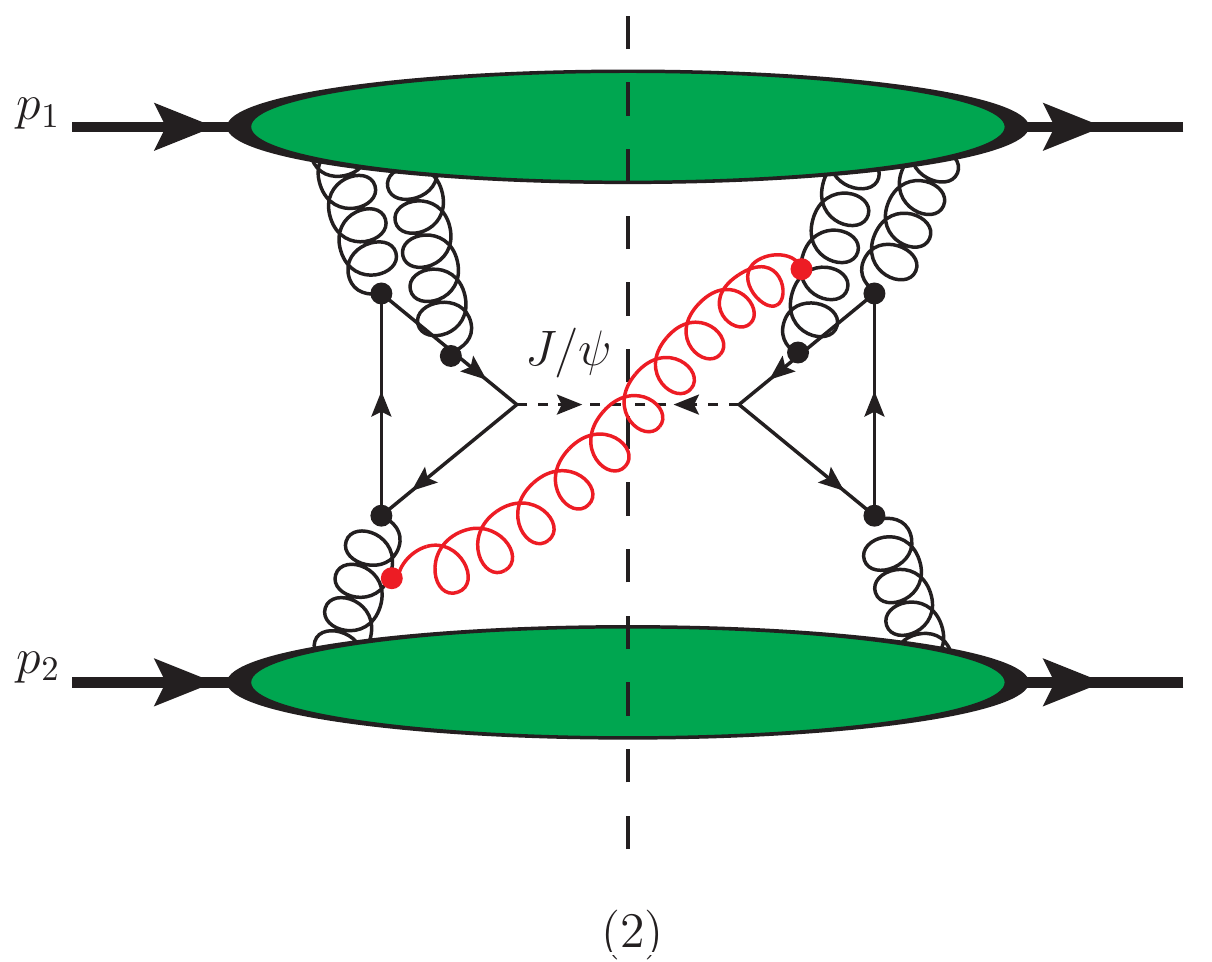}\includegraphics[scale=0.6]{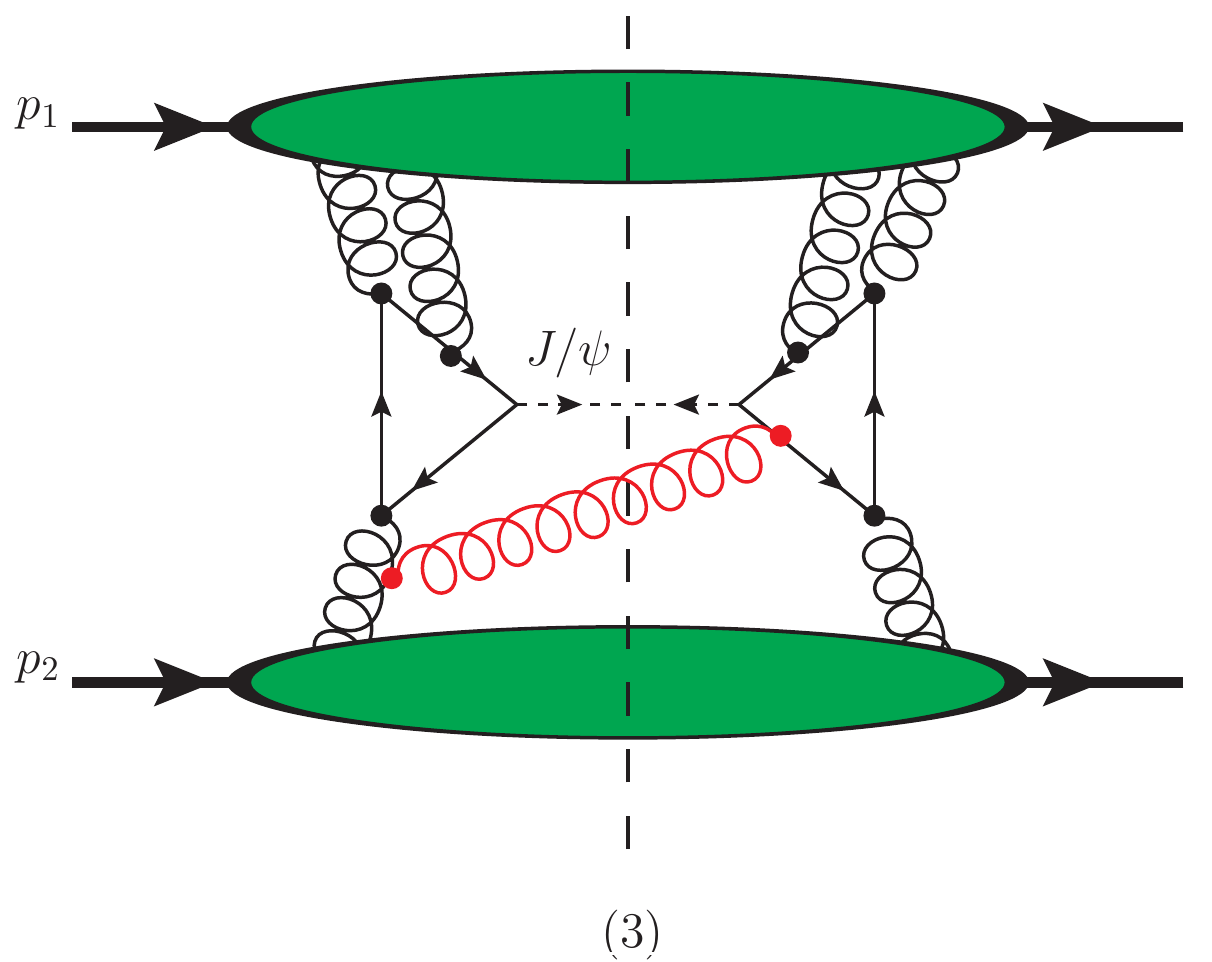}\caption{\label{fig:DiagExample}Diagram (1): Example of a diagram which contains
a double log and which after resummation contributes to gluon splitting
kernel $P_{gg}$. Another contribution to $P_{gg}$ comes from virtual
corrections (quark or gluon self-energy insertions into gluon lines),
as shown in the Diagram (1') . Diagrams 2 and 3: examples of diagrams
which possess collinear and soft divergencies. Though formally these
diagrams should be taken into account, as explained in the text, numerically
they give a very small contribution. In all three diagrams summation
over all permutations of gluons in quark loop is implied.}
\end{figure}

where $\sigma_{gg+g\to J/\psi\,g}$ is given by 
\begin{eqnarray}
d\sigma_{gg+g\to J/\psi\,g} & = & \frac{\alpha_{s}^{4}\left(\mu\right)}{8192\,\pi^{8}\,\hat{s}^{2}}\sum_{{\rm polarization}}\sum_{{\rm spin}}\sum_{{\rm color}}\left|\mathcal{M}_{gg+g\to g\,J/\psi}\right|^{2}\mathcal{F}\left(x_{1a},k_{1a\perp},x_{1b},k_{1b\perp},\Delta_{\perp}\right)\label{eq:dsigma}\\
 & \times & \mathcal{F}\left(x_{2},k_{2\perp}\right)\,d^{2}k_{1a\perp}d^{2}k_{1b\perp}d^{2}\Delta_{\perp}d^{2}k_{2\perp}dy\,d^{2}p_{T}\,dy_{g}\,dx_{1a}/x_{1a},\nonumber 
\end{eqnarray}
the unintegrated double gluon distribution $\mathcal{F}\left(x_{1a},k_{1a\perp},x_{1b},k_{1b\perp},\Delta_{\perp}\right)$
which appears in~(\ref{eq:dsigma}) is defined as~\cite{Diehl:2011yj,Gaunt:2009re}
\begin{eqnarray}
\mathcal{F}\left(x_{1a},\,k_{1a\perp},\,x_{1b},\,k_{1b\perp},\Delta_{\perp}\right) & = & \int d^{2}y_{\perp}\,e^{i\Delta_{\perp}\cdot y_{\perp}}\int\frac{dz_{1}^{-}}{2\pi}\frac{dz_{2}^{-}}{2\pi}\int d^{2}z_{1}^{\perp}d^{2}z_{2}^{\perp}e^{i\left(x_{1}z_{1}^{-}+x_{2}z_{2}^{-}\right)p^{+}}\\
 & \times & e^{-ik_{1}^{\perp}\cdot z_{1}^{\perp}-ik_{2}^{\perp}\cdot z_{2}^{\perp}}\left\langle p\left|\mathcal{O}_{a}\left(0,\,z_{1}\right)\mathcal{O}_{a}\left(y_{\perp},\,z_{2}\right)\right|p\right\rangle ,\nonumber \\
\mathcal{O}_{a}\left(y,\,z\right) & = & \Pi_{a}^{jj'}G^{+j'}\left(y-\frac{z}{2}\right)G^{+j}\left(y+\frac{z}{2}\right),
\end{eqnarray}
and the matrix $\Pi_{a}^{jj'}$ for gluon polarization labels $a=g,\,\Delta g,\,\delta g$
is given by 
\begin{eqnarray}
\Pi_{g}^{jj'} & = & \delta^{jj'},\quad\Pi_{\Delta g}^{jj'}=i\epsilon^{jj'},\quad\Pi_{\Delta g}^{jj'}=\tau^{jj',ll'},\\
\tau^{jj',ll'} & = & \frac{1}{2}\left(\delta^{jk}\delta^{j'k'}+\delta^{jk'}\delta^{j'k}-\delta^{jj'}\delta^{kk'}\right).\nonumber 
\end{eqnarray}
The diagram 3 in the Figure~\ref{fig:ProcessListCSM} is not gauge
covariant on its own and should be supplemented with additional diagrams
which contribute in the same order in $\mathcal{O}\left(\alpha_{s}(m_{c})\right)$,
as shown in the Figure~\ref{fig:Gauge}. The diagram 2 in the list
corresponds to a feed-down contribution to gluon uPDF from digluon
uPDF. The diagram 3 is a radiative correction to the process suggested
in~\cite{Motyka:2015kta} (diagram 2 in the Figure~\ref{fig:ProcessListCSM}).
The diagram $4$ gives nonzero contribution due to nontrivial color
structure of the gauge group: the color independent part of the diagram
with inverted direction of the quark loop contributes with opposite
sign, for this reason the sum yields a nonzero contribution 
\begin{equation}
\sim{\rm tr}\left(t^{a}t^{b}t^{c}t^{d}\right)-{\rm tr}\left(t^{d}t^{c}t^{b}t^{a}\right)=\frac{i}{8}\left(f_{abe}d_{cde}+f_{cde}d_{abe}\right)
\end{equation}
The evaluation is quite straightforward and was done with FeynCalc~\cite{Mertig:1990an,Shtabovenko:2016sxi}
package. An important technical observation which allows to simplify
significantly the evaluations is that the hard coefficient functions
of all the diagrams in the Figure~\ref{fig:Gauge} effectively reduce
to the sum over the permutations of four gluons in the $V_{gggg\to J/\psi}$
vertex, as shown in the Figure~\ref{fig:GluonVertex}. This allows
us to perform numerically the symmetrization instead of evaluating
explicitly all possible interference terms which stem from the amplitude
and in its conjugate. In numerical evaluations of particular concern
are the diagrams which stem from the interferences of the diagram
3 in the Figure~\ref{fig:Gauge}: when squared (see diagram 1 in
the Figure~\ref{fig:DiagExample}), in collinear limit they yield
(together with the virtual corrections shown in the diagram 1' of
the same Figure) the familiar gluon splitting kernel $P_{gg}$ \cite{Dokshitzer:1977sg,Gribov:1972ri,Altarelli:1977zs}.
When the diagram 3 interferes with other diagrams, as shown in diagrams
(2,~3) of the Figure~\ref{fig:DiagExample}, additionally it might
contain collinear and soft divergencies in certain points. Special
care is needed near the points where the different singularities start
overlapping and pinch the integration contour: in this case individual
diagrams might contain real singularities. Due to complex structure
of the integrand, demonstration of analytic cancellation of singularities
is challenging, for this reason we used a numerical method which will
be described in the section~\ref{sec:Results} below. Numerically
these diagrams give a very small contribution (see \emph{e.g}. the
Figure~\ref{fig:DiagExample}) and could be disregarded. This happens
because the average rapidities of the emitted gluons in the amplitude
and in its conjugate are different, and only a very small domain in
the configuration space contributes to the interference.

\begin{figure}
\includegraphics[scale=0.6]{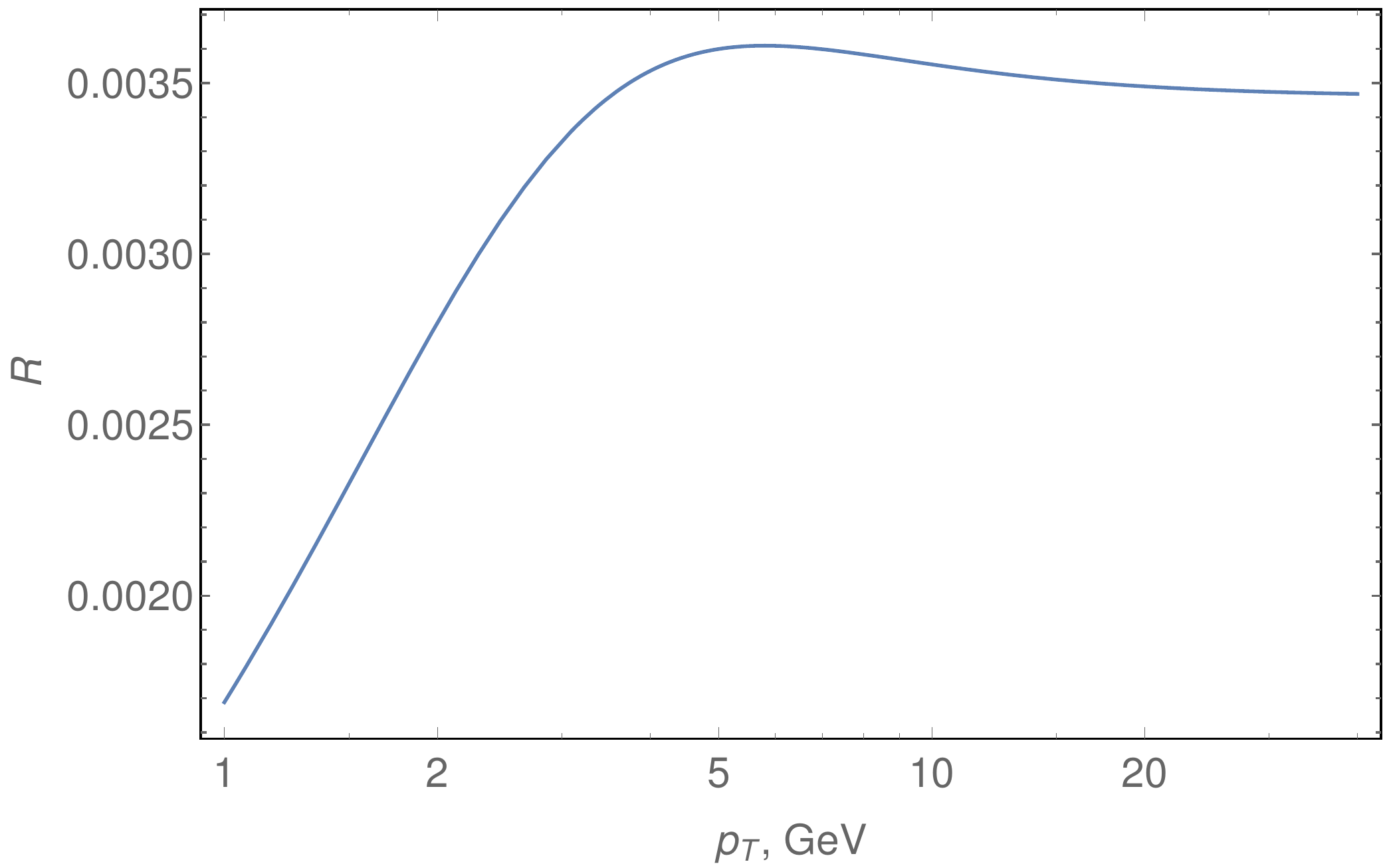}\\
 \caption{\label{fig:RelativeInterference}The relative contribution of the
diagram (2) from the Figure~\ref{fig:DiagExample} to the total result.}
\end{figure}

\section{Parametrization of gluon parton distributions}

\label{sec:Parametrization}For evaluation of the unintegrated gluon
parton densities $\mathcal{F}\left(x,\,k_{\perp}\right)$ we use Kimber-Martin-Ryskin
(KMR) parametrization~\cite{Kimber:2001sc} with collinear HERAPDF
NLO~\cite{Aaron:2009aa,Abramowicz:2015mha} gluon density used as
input. The color structure of the double gluon distribution in general
case is given by~\cite{Diehl:2011yj} 
\begin{eqnarray}
\mathcal{F}^{aa',bb'} & = & \frac{1}{64}\left[^{1}F\,\delta^{aa'}\delta^{bb'}-\frac{\sqrt{8}}{3}\,^{A}F\,f^{aa'c}f^{bb'c}+\frac{3\sqrt{8}}{5}\,^{S}F\,d^{aa'c}d^{bb'c}\right.\\
 &  & +\left.\frac{2}{\sqrt{10}}\,^{10}F\,\left(t_{10}^{aa',bb'}+\,\left(t_{10}^{aa',bb'}\right)^{*}\right)+\frac{4}{\sqrt{27}}\,^{27}F\,t_{27}^{aa',bb'}\right],\nonumber 
\end{eqnarray}
where $t_{i}$ are generators of the color group in representation~($i=10,\,\bar{10},\,27$).
In what follows, for the sake of simplicity we will consider that
the color structure is given by only the first term $\sim\delta^{aa'}\delta^{cc'}$,
tacitly omitting other terms with nontrivial color structure. This
choice does not violate any of the positivity bounds mentioned in~\cite{Diehl:2013mla}.
For the kinematic dependent terms, we assume that emission of both
gluons is uncorrelated and use the model suggested in~\cite{Diehl:2011yj}
with additional $k_{T}$-dependence, 
\begin{equation}
\mathcal{F}\left(x_{1a},\,k_{1a\perp},x_{1b},\,k_{1b\perp},\Delta_{\perp}\right)=\mathcal{F}\left(x_{1a},\,k_{1a\perp}\right)\mathcal{F}\left(x_{1b},\,k_{1b\perp}\right)e^{-B_{g}\Delta_{\perp}^{2}}\label{eq:F_param}
\end{equation}
where the value of the diffractive slope $B_{g}$ is taken as a sum
of values of gluon GPD slope~\cite{Goloskokov:2006hr}, 
\begin{equation}
B_{g}\approx\left(2\times2.58+0.15\ln\left(1/x_{1a}\right)+0.15\ln\left(1/x_{1b}\right)\right)\,{\rm GeV^{-2}}.
\end{equation}
For the case of the double parton scattering process $pp\to h_{1}h_{2}X$,
this parametrization leads to the so-called ``pocket formula'' 
\begin{equation}
d\sigma_{pp\to h_{1}h_{2}X}=\frac{d\sigma_{pp\to h_{1}X}\,d\sigma_{pp\to h_{2}X}}{\sigma_{eff}}
\end{equation}
where the cross-section $\sigma_{eff}$ is a functional of the impact
parameter profile of the parton distribution~\cite{Diehl:2011yj}.
The experimental estimates of $\sigma_{eff}$ from DPS depend on the
hadrons $h_{1},\,h_{2}$ with typical values $\sigma_{eff}\approx6-15\,$mb~\cite{Aad:2013bjm,Aaboud:2016fzt,Shao:2016wor}.

In the forward limit ($\Delta_{\perp}\to0$), which is much better
understood due to smaller number of variables, after integration over
the transverse momenta $k_{i\perp}$, the parametrization~(\ref{eq:F_param})
yields for the collinear digluon distributions 
\begin{equation}
G\left(x_{1},\,x_{2},\,\mu_{F}^{2}\right)=G\left(x_{1},\,\mu_{F}^{2}\right)G\left(x_{2},\,\mu_{F}^{2}\right)\label{eq:g_fact}
\end{equation}
Recently in~\cite{GolecBiernat:2015aza} it was suggested a model
of collinear digluon densities which takes into account all known
sum rules and evolution equations. While in general their result might
differ quite significantly from a factorized form~(\ref{eq:g_fact}),
for large factorization scale $\mu_{F}^{2}\gtrsim M_{J/\psi}^{2}$
and small $x_{1,2}\ll1$ the factorized form~(\ref{eq:g_fact}) holds
within 10\%. This result agrees with more general result of~\cite{Diehl:2014vaa}
that evolution to higher scales relevant for quarkonia production
tends to wash out any correlations present at low scales.

\section{Numerical results}

\label{sec:Results}
\begin{figure}
\includegraphics[scale=0.7]{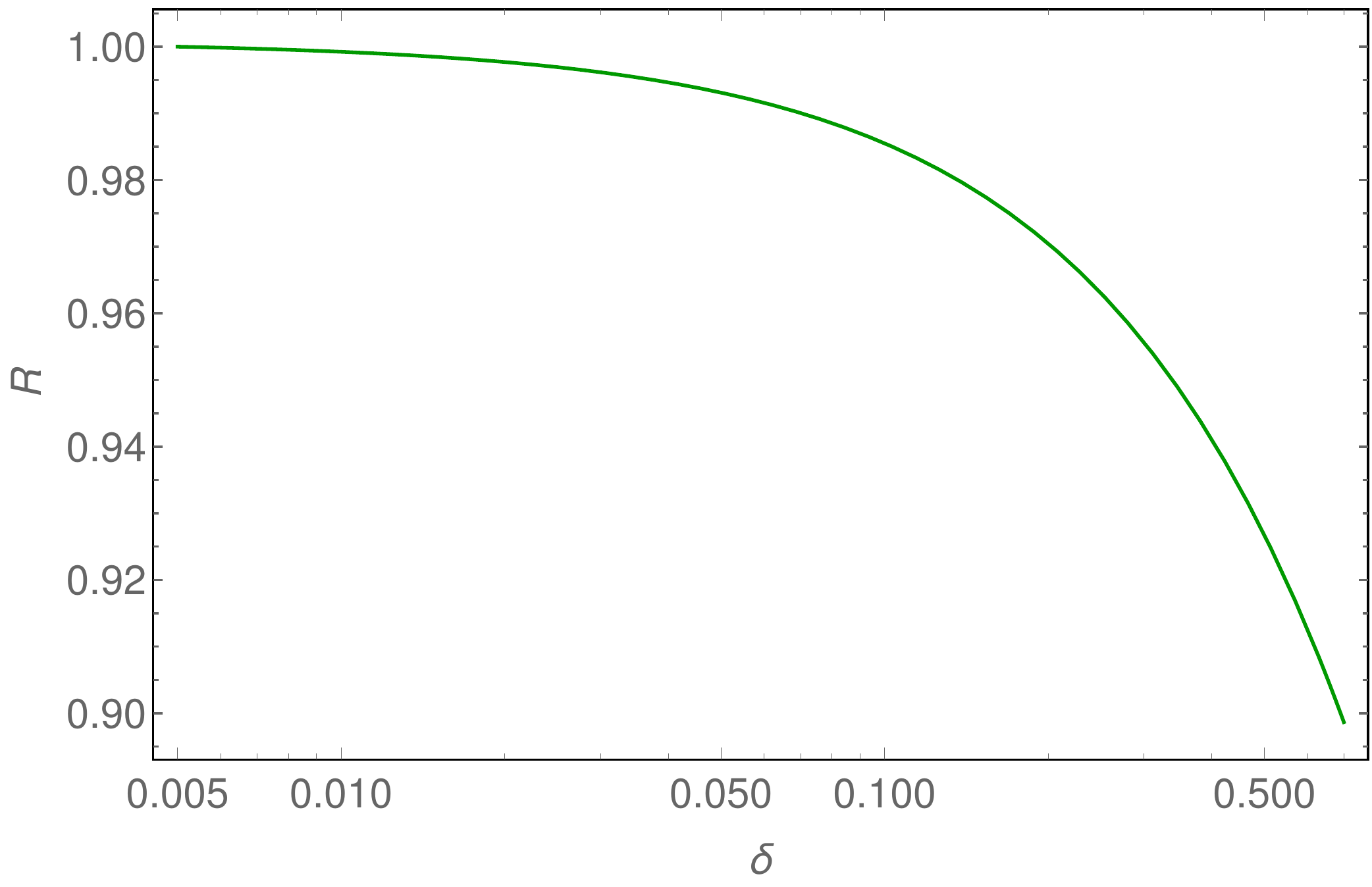}\caption{\label{fig:delta_dep}Dependence of the ratio $R$ defined in~(\ref{eq:R_def})
on parameter $\delta$. Stability of the result at small $\delta$
is a numerical manifestation that collinear divergencies cancel in
the full sum.}
\end{figure}

As was discussed in Section~\ref{sec:Evaluation},
the amplitude might contain soft and collinear divergencies in certain
points. It is quite challenging to demonstrate analytically that such
cancellation indeed happens, for this reason we use a numerical method
suggested in~\cite{Yuasa:2011ff,deDoncker:2012qk} and implemented
in SecDec package~\cite{Carter:2010hi,Borowka:2012yc,Borowka:2015mxa}
widely used for numerical multiloop evaluations. This method consists
in treating the Feynman regularizer $+i\delta$ as a finite parameter,
\begin{equation}
S(p)=\frac{\hat{p}-m}{p^{2}-m^{2}+i\delta}.
\end{equation}
Similarly, we treat $+i\delta$ as a finite parameter in the gluon
propagator complemented with Mandelstam-Leibbrandt prescription~\cite{Mandelstam:1982cb,Leibbrandt:1987qv}
\begin{equation}
\frac{1}{k^{+}}\to\frac{k^{-}}{k^{+}k^{-}+i\delta}.
\end{equation}
As was discussed in~\cite{Yuasa:2011ff,deDoncker:2012qk}, the infrared
and collinear singularities in individual diagrams translate into
poles in $\delta$, which however should eventually cancel in the
infrared stable result. In the Figure~\ref{fig:delta_dep} we plot
the ratio 
\begin{equation}
R(\delta)=\frac{d\sigma\left(\delta\right)}{d\sigma\left(5\times10^{-3}\right)}\label{eq:R_def}
\end{equation}
as a function of parameter $\delta$. Stability of the result for
small $\delta$ ensures that the result is free of any infrared divergencies.

\begin{figure}
\includegraphics[scale=0.5]{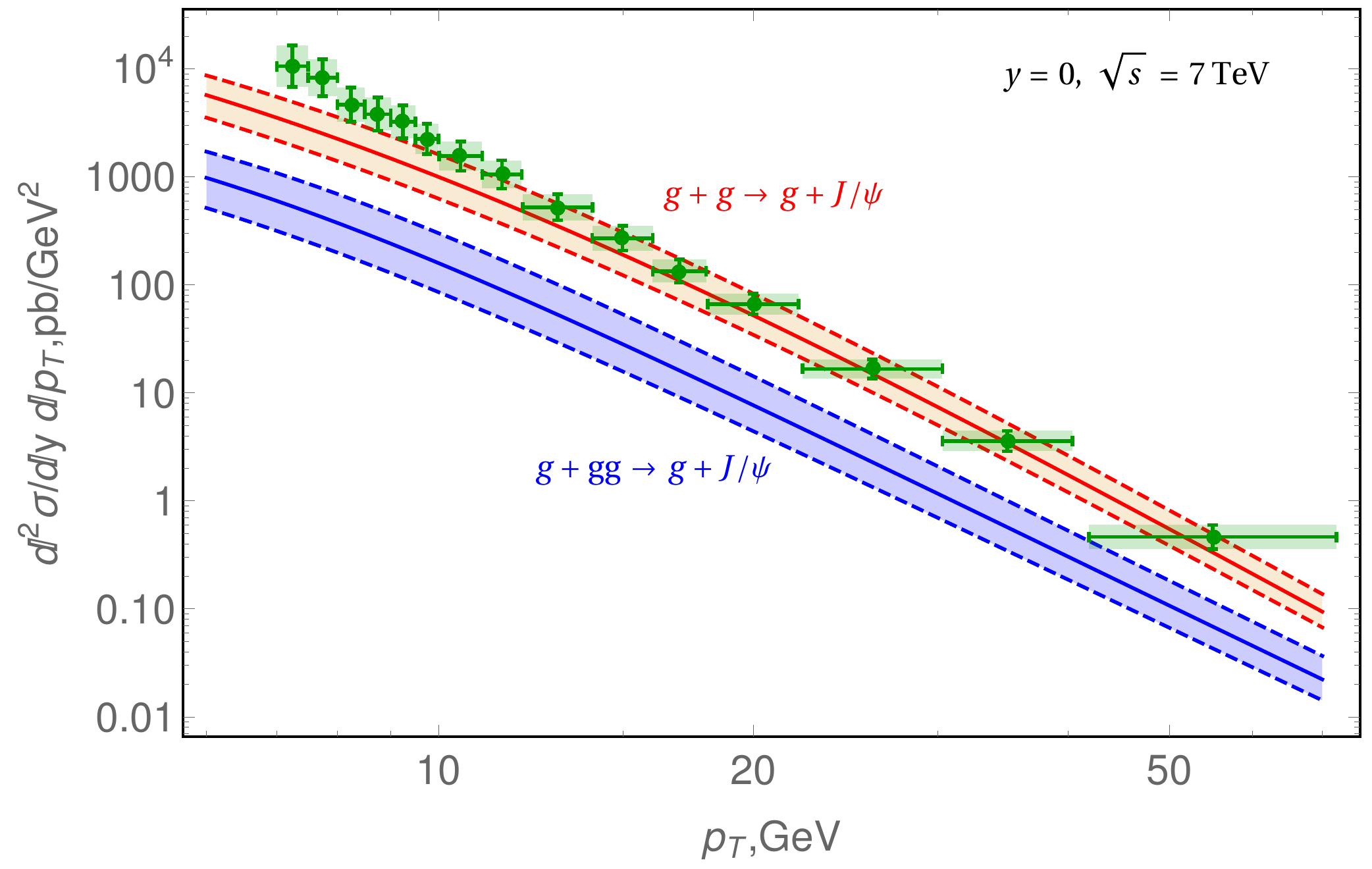}\\
 \includegraphics[scale=0.5]{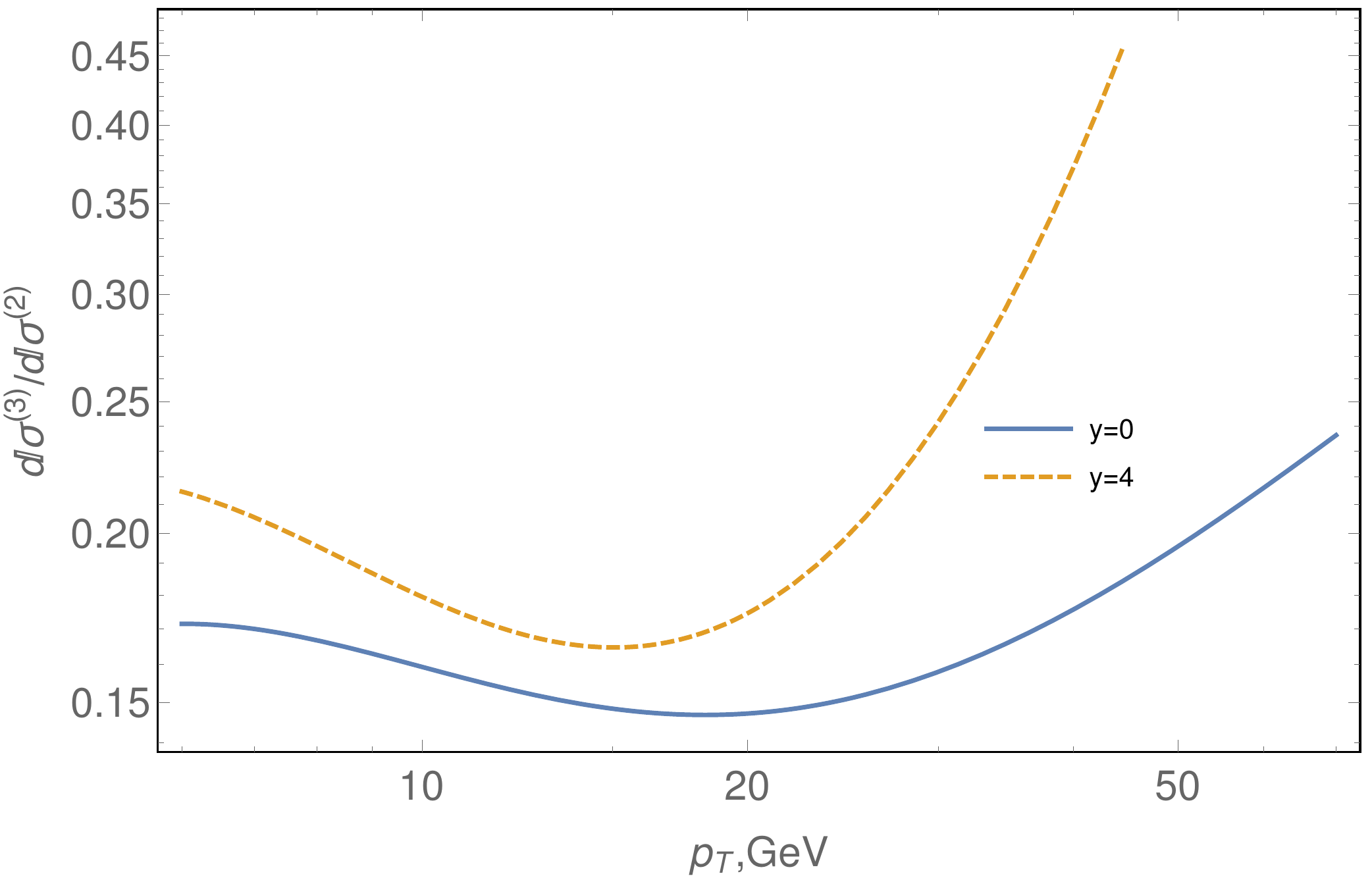}\caption{\label{fig:Results}(color online) Top: Cross-section of prompt $J/\psi$
production (sum of direct and feed-down contributions) evaluated in
CSM framework (upper red band) and digluon correction (lower blue
band). The errorbars illustrate uncertainty due to higher order loop
corrections and are estimated varying the renormalization scale $\mu_{R}$
in the range $\mu_{R}\in\left(0.5,\,2\right)\times\sqrt{p_{\perp}^{2}+M_{J/\psi}^{2}}$.
Experimental points (green boxes) are from ATLAS~\cite{Aad:2011sp}.
Bottom: Ratio of our mechanism to the Color Singlet Mechanism as a
function of $J/\psi$ transverse momentum $p_{T}$. }
\end{figure}

In the Figure~\ref{fig:Results} we compare contribution of our mechanism
with $k_{T}$-CSM results for prompt $J/\psi$ production . As we
can see, the contribution is enhanced at large $p_{T}$ and for $p_{T}\gtrsim50$
GeV at forward rapidities presents a sizeable contribution to the
total result. However, in the $p_{T}$-integrated cross-section, which
is dominated by small-$p_{T}$ domain, the considered contribution
is small ($\lesssim20$ per cent even at forward rapidities), and
by the order of magnitude agrees with mechanism~\cite{Motyka:2015kta}.
For the sake of definiteness, we fixed the renormalization and factorization
scales as $\mu_{R}=\mu_{F}=\sqrt{p_{\perp}^{2}+M_{J/\psi}^{2}}$ and
estimate the higher order loop corrections varying the scale in the
range $(0.5,\,2)\sqrt{p_{T}^{2}+M_{J/\psi}^{2}}$, in agreement with~\cite{Brodsky:1982gc}.
However, we would like to mention that for three-gluon vertex this
prescription might be not very accurate since the effective scale
in this case is controlled by the smallest virtuality~\cite{Binger:2006sj}
(which means that loop corrections could be large).

\section{Conclusions}

\label{sec:Conclusions}In this paper we studied the contribution
of the double parton gluon densities to the $J/\psi$ production.
Though formally suppressed in the heavy quark mass limit, the suggested
mechanism is significant and constitutes up to twenty per cent of
the produced $J/\psi$, on par with the the contribution suggested
in~\cite{Motyka:2015kta}. The suggested mechanism is not suppressed
at large quarkonia momenta $p_{T}$, and for this reason presents
one of the possible mechanisms of charmonia production in this kinematics. 

The considered contribution grows with energy, and we expect that
similar trend holds for higher order multigluon contributions. At
sufficiently small $x_{B}$, eventually we approach the saturation
regime, which is usually described by the phenomenological small-$x_{B}$
models with built-in saturation, like dipole model~\cite{Kowalski:2003hm,Rezaeian:2012ji,gbw01:1,Bartels:2002cj}
or CGC~\cite{McLerran:1993ni,McLerran:1993ka}. These models can
describe the $J/\psi$ production~\cite{Kopeliovich:2017jpy,Kang:2013hta,Fujii:2005pg},
however the relation of the nonperturbative dipole cross-section to
single and multiple gluon distributions in the DGLAP framework might
be not straightforward and rely on model-dependent assumptions~\cite{Altinoluk:2014oxa,Kowalski:2003hm,Rezaeian:2012ji,gbw01:1,Bartels:2002cj}.
In case when the model admits interpretation in terms of the gluon
distributions, usually the multigluon distributions are hard-coded
in the underlying model, frequently being a simple product of single-gluon
uPDFs in the impact parameter space~\cite{Rezaeian:2012ji}. At the
same time, recent theoretical~\cite{Diehl:2011yj,GolecBiernat:2015aza,Rinaldi:2013vpa,Rinaldi:2016mlk,Diehl:2014vaa}
and experimental~\cite{Khachatryan:2014iia,Aaboud:2016dea,Aaboud:2016fzt,Aad:2011sp,Aad:2013bjm,Abazov:2014fha,Abazov:2015nnn,Abe:1997bp,Chatrchyan:2013xxa}
studies suggest that gluon DPDFs might be much more complicated objects
due to possible correlation between partons~\cite{Calucci:2010wg},
and in view of various sum rules which the DPDFs should satisfy~\cite{GolecBiernat:2015aza}.
In contrast to the small-$x$ models, the suggested approach does
not use eikonal approximation and can be used with arbitrary gluon
DPDFs extracted from DPS experiments.

\section*{Acknowldgements}

We thank our colleagues at UTFSM university for encouraging discussions.
Our special thanks go to Stanley Brodsky, who suggested the topic
of this research and participated in some discussions. We also thank
Sergey Baranov for discussions and technical clarifications regarding
the references~~\cite{Baranov:2002cf,Baranov:2007dw}. This work
was supported in part by Fondecyt (Chile) grants 1140390 and 1140377,
by Proyecto Basal FB 0821 (Chile), and by CONICYT grant PIA ACT1406
(Chile). Powered@NLHPC: This research was partially supported by the
supercomputing infrastructure of the NLHPC (ECM-02). Also, we thank
Yuri Ivanov for technical support of the USM HPC cluster where a part
of evaluations has been done. 

 \end{document}